\documentstyle[12pt]{article}
\begin{document}
\begin{center}
{\large \bf Has superluminal light propagation been observed?
         \footnote{ The project supported  partially by the Ministry of
          Science and Technology of China under Grant No. 95-Yu-34 and
          National Natural Science Foundation  
          of China under Grant Nos. 19745008 and 19835040} }\\

\vspace{6mm}
Yuan-Zhong Zhang\footnote{  Email: yzhang@itp.ac.cn }
 
\vspace{4mm}
{\footnotesize\it CCAST (World Laboratory), P.O. Box 8730, Beijing, China\\
Institute of Theoretical Physics, Chinese Academy of Sciences, \\
           P.O. Box 2735, Beijing, China} \footnote{Mailing address }\\
\end{center}
\vspace{5mm}

\noindent
{\bf It says in the report$^1$  by Wang et al. that a negative group velocity
$u=-c/310$ is obtained and that a pulse advancement shift 62-ns is measured.
The authors claim that the negative group velocity is associated with
superluminal light propagation and that the pulse advancement is not at odds
with causality or special relativity. However, it is shown here that their
conclusions above are not true. Furthermore, I give some suggestion
concerning a re-definition of group-velocity and a new explanation in special
relativity of causality.}  

{\bf The velocity of ${\bf u}=-(c/310){\bf \hat{k}_0}$ is subluminal but not
superluminal} (the term 
``superluminal" is usually understood as such a light propagation with phase,
group, and energy velocities all exceeding the vacuum speed of light$^2$).  
It is well-known that the 4-dimensional interval for a signal in special
relativity is given by 
\begin{equation}
   ds^2 =c^2 dt^2  -\left(dx^2 +dy^2 +dz^2\right) =dt^2 \left(c^2 -u^2\right),
\end{equation}
as seen in the inertial frame $K$, where
\begin{equation}
   u^2 ={\bf u}\cdot{\bf u}=\left(\frac{dx}{dt}\right)^2 +
     \left(\frac{dy}{dt}\right)^2 + \left(\frac{dz}{dt}\right)^2  
\end{equation}
with ${\bf u}$ being the velocity of the signal. According to special
relativity, $ds^2$ is an invariant under Lorentz transformations, i.e. $ds^2
= {ds'}^2$. This means  
\begin{equation}
   {dt'}^2 \left(c^2 -{u'}^2\right)=dt^2 \left(c^2 -u^2\right),
\end{equation} 
where the quantities with a prime stand for the ones as seen in other
inertial frame $K'$. This shows that both $u^2$ and $u'^2$ are all bigger, or
all less, than $c^2$. Explicitly, $|{\bf u}|<c$ leads to $|{\bf u'}|<c$.
Similarly,  $|{\bf u}|>c$ leads to $|{\bf u'}|>c$. For the case in the report
by Wang et al., the velocity is found to be (in terms of vector symbol) 
${\bf u}=-(c/310){\bf \hat{k}_0}$ with ${\bf \hat{k}_0}$ being the unit
vector of the incident direction (see below) and hence $|{\bf u}|<c$ as seen
in the laboratory frame. So that we 
have $|{\bf u'}|<c$, i.e. the velocity of the pulse would also be smaller
than the vacuum speed of light $c$, as seen in any  of other inertial frames.
Therefore, the negative velocity obtained by Wang et al. is simply subluminal
but not superluminal.  

{\bf New suggestion concerning re-definition of negative velocity.}
Now I want to give a new explanation of the so-called ``negative" group
velocity. By definition the group velocity of a light pulse propagating in a
dispersive linear medium is given by$^3$ 
\begin{equation}
   {\bf u}=\frac{\partial \omega}{\partial \bf k}=\frac{c}{n+\nu\frac{{\rm
            d}n}{{\rm d}\nu}}{\bf \hat{k}},   
\end{equation}
where ${\bf\hat{k}}\equiv {\bf k}/|{\bf k}|$ is the unit vector of the
direction of phase velocity 
(or wave vector), ${\bf k}$ is wave vector, $\nu =\omega/2\pi$ is
frequency, and $n=n(\nu)$ is the optical refractive index of the medium.  

For a normal medium we have ${\rm d}n/{\rm d}\nu >0$ and so that 
$|{\bf u}|<c$. But for anomalous dispersive linear media in where  
${\rm d}n/{\rm d}\nu<0$ , one arrives at the following two situations: 
(i) For $1> n(\nu)+\nu {\rm d}n/{\rm d}\nu \geq 0$, we have $|{\bf u}|>c$; 
(ii) For $n(\nu)+\nu {\rm d}n/{\rm d}\nu <0$, one gets 
\begin{equation}
   {\bf u}=-\frac{c}{|n+\nu\frac{dn}{d\nu}|}{\bf\hat{k}}.  
\end{equation}
In case of a light pulse propagating vertically towards a surface of
dispersive medium from vacuum, the incident direction ${\bf \hat{k}_0}$ is
usually defined as to be positive. The wave vector (or phase velocity) of the
pulse in the medium is usually assumed to have a positive direction (i.e. 
${\bf\hat{k}}={\bf\hat{k}_0}$) while the group
velocity ${\bf u}$ then has a negative direction$^4$. If ${\bf u}$ represents
the velocity of an actual information, then the definition of negative
group-velocity must give violation of causality. Thus it is needed to modify
the usual definition of phase and group velocities.  

Here it must be emphasized that the negative sign ``$-$" in Eq. (5) simply
indicates the directions of the group-velocity ${\bf u}$ and wave-vector
${\bf k}$ are opposite each other, but not say which one should be negative.
In fact there is no reason to identify the direction of ${\bf k}$ in the
medium with the incident one ${\bf\hat{k}_0}$. Contrarily,  it should be more
reasonable to 
suppose the group-velocity ${\bf u}$ has the same direction to that of the
incident light signal, while the wave-vector ${\bf k}$ ( and hence phase
velocity) then has a negative direction, i.e., ${\bf\hat{k}}=-{\bf\hat{k}_0}$. 
By use of the new definition, we
never meet any problem concerning violation of causality in case where the
group-velocity {\bf u} does represent a velocity of an actual information. 

Now come back to the case of $u=-c/310$ in terms of the symbol by Wang et al.. 
Note that the negative velocity is
not directly measured but calculated by Wang et al. from their measured
refraction index by use of the definition: $u=c/n_{g}$ with $n_{g}\equiv
n(\nu)+\nu {\rm d}n(\nu)/{\rm d}\nu$. In other words, 
the negative sign ``$-$" for $n_{g}<0$ is just defined by them. Contrarily,
according to the present new definition, the group-velocity calculated from
their measured refraction index should be positive, and less than $c$.  

{\bf The observed 62-ns advancement shift must be violation of causality.}
Another result in the report$^1$ by Wang et al. is the 62-ns advancement
shift (see Fig. 4 of Ref. 1). They claim that it is not at odds with
causality. They argue that it is a result of the wave nature of light and
that no actual information, or signal, is trasmitted$^{1, 5}$. However it
must be pointed out that the authors make confusion of the direct observation
with theoretical prediction. At first it is emphasized that the 62-ns shift
is a directly measured datum but not a theoretical prediction. Secondly it is
needed to clear whether the observed 62-ns shift is an actual information. If
not, one must face the question: Can you measure a non-actual information in
a laboratory? In fact, it is not possible for any experimental device to
record a non-actual signal. In other words, what a device records is
certainly an actual information. So that the curves A and B in Fig. 4 in the
report$^1$ are just the records of actual information.  
The curves A and B show that the actual signal B is advanced for 62-ns in
time compared to the the actual signal A. If A were the source of B, then  
the 62-ns advancement would certainly be violation of causality. This
conclusion is independent of any theoretical prediction concerning phase,
group, or other kind of velocity. Owing to any actual signal should not
violate causality, then the curve B could be connected causally not with the
curve A but only with a measurement error. The 3.7-$\mu$s full-width at
half-maximum of the probe pulse   
means the pulse spatial extension of more than 1-km much larger than the 6-cm
length of the atomic cell. On the other hand, the curve B is only translated
in time but almost not changed in shape compared to the curve A. 
So that possible sources of  the 62-ns translation would be a systematic
error, or a pulse-reshaping phenomenon such as the amplification of the
pulsefront and reduction of its tail.  In order to determine finally the
source of the advancement shift, it is needed to perform further similar
measurements in different experiment conditions, such as different probe
pulses, different cell lengths, and so on. 

{\bf New suggestion for explanation of causality in special relativity.} 
For an anomalous dispersive medium with $ 1> n(\nu)+\nu {\rm d}n/{\rm d}\nu
\geq 0$, group-velocity is superluminal (i.e. $u>c$) in laboratory frame in
which the medium is at rest. Eq. (3) gives that the group-velocity in any of
inertial frames is still superluminal  (i.e. $u'>c$).   
Let $t_1$ and $t_2$ be the instants at which the light signal arrives at    
 points 1 and 2, respectively, in the medium.  Due to the fact of $\triangle
t=t_2 -t_1 >0$,  no causality would be violated in the laboratory frame. But
by making use of Lorentz transformations, one always find such an inertial
frame, e.g. the frame $K'$, in which we have $\triangle t' =t'_{2}-t'_{1}<0$.
This is just the so-called violation of causality as seen within $K'$.
However it must be addressed that $\triangle t^{'}$ and $\triangle t$ are
coordinate time intervals but not proper ones. It is well-known that a
coordinate time interval is related to the definition of simultaneity and
thus is not directly observable$^6$. In special relativity, therefore, all of
physical observations must be used to compare with such quantities which are
invariant under Lorentz transformations, while the only exception is just the
explanation of causality above. Here I suggest to explain causality by means
of proper time interval in stead of coordinate one. To do it, let the signal
come back to the point 1 after it reaches the point 2. In this case we have  
$\triangle \tau =t_3 -t_1 >0$ where $t_3$ is the instant at which the signal
returns to the point 1. Due to $t_1$ and $t_3$ are readings of the same
standard clock at rest at the point 1, so that $\triangle \tau$ is just a
proper time interval to be positive in all of inertial frames. Using the new
definition one could arrive at the conclusion: The superluminal light
propagation (i.e. $u>c$) is not at odds with both causality and special
relativity.

\end{document}